\documentclass[3p,times,procedia]{elsarticle}
\usepackage{nupha_ecrc}

\RequirePackage{lineno}

\volume{00}

\firstpage{1}

\journalname{Nuclear Physics A}

\runauth{}

\jid{nupha}

\jnltitlelogo{Nuclear Physics A}

\usepackage{amssymb}
\usepackage[figuresright]{rotating}

\usepackage{natbib}

\begin{document}

\begin{frontmatter}



%% Instructions from Editor: Please use the following \dochead only in the preprint version (e-print arXiv etc.); 
%% use empty \dochead{} when submitting to Nuclear Physics A!
\dochead{XXVIth International Conference on Ultrarelativistic Nucleus-Nucleus Collisions\\ (Quark Matter 2017)}
%% e.g. \dochead{17th International Conference on Dynamical Processes in Excited States of Solids}

\title{Disentangling flow and signals of Chiral Magnetic Effect in U+U, Au+Au and p+Au collisions}


\author{Prithwish Tribedy (for the STAR Collaboration)}%~\footnote{A list of members of the STAR Collaboration and acknowledgements can be found at the end of this issue.}}

\address{Physics Department, Brookhaven National Laboratory, Upton, NY 11973, USA}

\begin{abstract}
We present STAR measurements of the charge-dependent three-particle correlator $\gamma^{a,b}=\langle \cos(\phi_1^{a}+\phi_2^{b}-2\phi_3)\rangle/v_{2}\{2\}$ and elliptic flow $v_{2}\{2\}$ in U+U, Au+Au and p+Au collisions at RHIC. The difference $\Delta \gamma = \gamma(\rm{opposite\!-\!sign})\!-\!\gamma(\rm{same\!-\!sign})$ measures charge separation across the reaction plane, a predicted signal of the Chiral Magnetic Effect (CME). Although charge separation has been observed, it has been argued that the measured separation can also be explained by elliptic flow related backgrounds. In order to separate the two effects, we perform measurements of the $\gamma$-correlator where background expectations differ from magnetic field driven effects. A differential measurement of $\gamma$ with the relative pseudorapidity ($\Delta\eta$) between the first and second particles indicate that $\Delta \gamma$ in peripheral A+A and p+A collisions are dominated by short-range correlations in $\Delta\eta$. However, a relatively wider component of the correlation in $\Delta\eta$ tends to vanish the same way as projected magnetic field as predicted by MC-Glauber simulations.

\end{abstract}


\end{frontmatter}


%\linenumbers

\section{Introduction}
\label{introduction}
QCD anomaly driven chirality imbalance can introduce an electromagnetic (vector) current along the direction of strong magnetic fields ($\vec{B}$) produced in relativistic heavy-ion collisions (HIC)~\cite{Kharzeev:1998kz,Kharzeev:2004ey,Kharzeev:2007jp,Mueller:2016ven}. 
Such a phenomenon, referred to as the Chiral Magnetic Effect (CME), is expected to lead to separation of produced charged hadrons in HICs along $\vec{B}$~\cite{Kharzeev:2004ey}. However, since the direction of $\vec{B}$ is not known in HICs, the angular correlations of a pair of charged particles with respect to the reaction plane $\Psi_{RP}$, given by $\gamma=\langle \cos(\phi_1+\phi_2-2\Psi_{RP})\rangle$ was proposed to be an experimental observable of the CME~\cite{Voloshin:2004vk}. The most common proxy for $\Psi_{RP}$ is the second harmonic event plane $\Psi_{2}$ corresponding to the elliptic flow coefficient $v_{2}$ of inclusive charged particles. Therefore, experimental searches for the CME concentrate on the effects of charge separation driven by the component of $\vec{B}$ along  $\Psi_{2}$~\cite{Abelev:2009ac,Adamczyk:2014mzf}. Any non-CME phenomenon that can lead to charge-dependent azimuthal correlations with respect to $\Psi_{2}$ can also influence $\gamma$~\cite{Voloshin:2004vk}. The goal of this analysis is to disentangle the effects of such potential non-CME backgrounds from the expected signals of CME.

Known sources of flow-driven backgrounds are HBT, Coulomb, flowing resonances, local charge conservation and momentum conservation~\cite{Voloshin:2004vk,Schlichting:2010qia,Wang:2009kd,Bzdak:2010fd}. 
A naive expectation of such flow-driven background contributions can be shown to be proportional to $v_{2}/N$, $N$ being the multiplicity~\cite{Voloshin:2004vk}. Attempts to reduce flow-driven background by reducing $v_{2}$ also reduces the ability to resolve the direction of $\Psi_2$ and therefore the effect of $\vec{B}$, resulting in a reduction of expected signal. It turns out that the centrality dependence of $v_{2}$ and the component of $\vec{B}$ along $\Psi_2$ are very similar making the signal-background separation challenging in HICs. However, there is an important distinction, unlike event-averaged $v_2/N$, MC-Glauber simulations indicate that the component of $\vec{B}$ along $\Psi_2$ can vanish due to decorrelation effects in the ultra-central A+A, peripheral A+A and in p+A collisions~\cite{Chatterjee:2014sea,Bloczynski:2013mca,Belmont:2016oqp}. In such scenarios, a naive background expectation (non-zero $v_2/N$) for charge separation differs from the vanishing $\vec{B}$-field driven expectation and provides a way to disentangle the two effects. 
In addition, non-flow effects such as fragmentation of (mini-)jets, particularly dominant in peripheral events, that can influence the orientation of $\Psi_2$ are also expected to contribute to the background~\cite{Abelev:2009ac}. Differential measurements of three-particle correlations with relative pseudorapidity in peripheral A+A and p+A collisions will be important to understand such effects~\cite{Khachatryan:2016got}. 

\section{Experiment and analysis}

\label{expt}

We analyzed the data on U+U at $\sqrt{s_{_{NN}}}=193$ GeV and Au+Au, p+Au collisions at $\sqrt{s_{_{NN}}}=200$ GeV respectively as collected by the STAR detector~\cite{Ackermann2003624} during 2011, 2012 and 2015 years of running of RHIC. For the measurements of $\gamma$ we used charged particles within the pseudorapidity range of $|\eta|\!<\!1$ and transverse momentum of $p_T\!>\!0.2$ GeV/$c$ detected by the Time Projection Chamber (TPC), the primary tracking systems of STAR situated inside a 0.5 Tesla solenoidal magnetic field~\cite{Anderson2003659}. We estimate the quantity $\gamma^{a,b}=C_{112}/v_{2}\{2\}=\langle \cos(\phi_1^{a}+\phi_2^{b}-2\phi_3)\rangle/v_{2}\{2\}$ with different charge combinations $a,b=+-,++,--$ and $v_{2}\{2\}^{2}=\langle\cos(2(\phi_1-\phi_2))\rangle$ as a measure of the elliptic flow coefficient obtained using algebra based on Q-vectors. In order to account for imperfections in the detector acceptance, we apply track-by-track weighting~\cite{Bilandzic:2010jr, Bilandzic:2013kga}. We also apply momentum-dependent tracking efficiency corrections. We estimate systematic uncertainties in our measurements by analyzing datasets with different efficiency estimates, by varying z-vertex position of the collision, and by varying track selection criteria. We estimate the number of participant nucleons $N_{\rm part}$ using a Monte-Carlo Glauber model for different centrality intervals. For the selection of such centrality bins we use the distribution of minimum bias uncorrected multiplicity of charged particles in the pseudorapidity region $|\eta|<0.5$ measured by the TPC and using the response of spectator neutrons in the Zero-Degree-Calorimeters (ZDCs).

\section{Results and discussion}
\label{results}

 \begin{figure}[t]
 \includegraphics[width=1\textwidth]{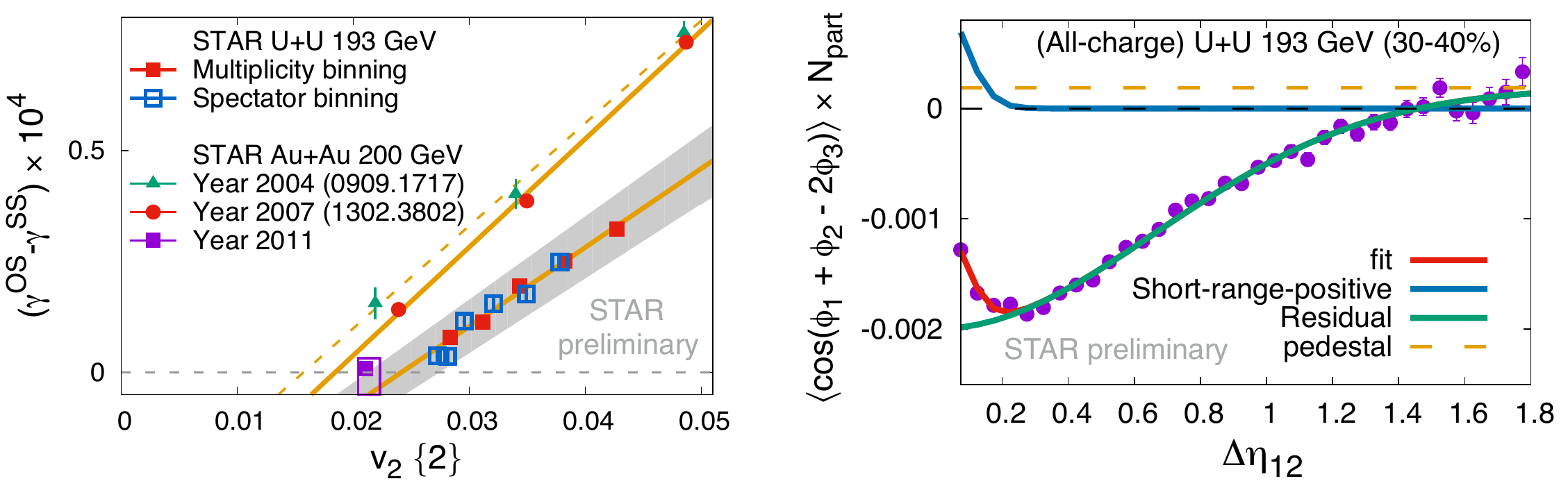}
%\includegraphics[width=0.5\textwidth]{dgamma_vs_v2.eps}
%\includegraphics[width=0.5\textwidth]{c112_all_30_40.eps}
%\hspace{2pc}%
\caption{\small \label{gammav2} (left) The variation of $\Delta\gamma$ with $v_{2}$. (right) Relative pseudorapidity dependence of the three-particle correlator (numerator of $\gamma$) shown for all charges, the curves show three component fit to the data points.}
\end{figure}

Fig.\ref{gammav2} (left) shows the correlation of charge separation ($\Delta \gamma = \gamma^{{opposite\!-\!sign}}\!-\!\gamma^{{same\!-\!sign}}$) with $v_2\{2\}$ in central events in Au+Au (0-20$\%$) and U+U (0-10$\%$) collisions. Each point in this plot is obtained by first binning the distribution of multiplicity (Au+Au, U+U) or spectators (U+U) and then estimating the values of $v_2$ and $\Delta \gamma$ separately for the corresponding event classes. Interestingly, for both Au+Au and U+U, $\Delta \gamma$ trends to vanish for non-zero values of $v_2\{2\}$. 
The dependence of $v_2\{2\}$ {\it vs.} $\Delta \gamma$ in background model ($\propto v_2\{2\}/N$) is non-linear since $N$ increases with decreasing $v_2\{2\}$ towards central events. It will therefore be interesting to see if a background calculation like Ref~\cite{Ma:2011uma} can predict $\Delta \gamma$ {\it vs.} $v_{2}\{2\}$ with a positive intercept on $v_{2}\{2\}$-axis. 
 A simultaneous description of the vanishing $\Delta \gamma$ at non-zero $v_2\{2\}$ and its rapid linear growth can be naturally explained by the variation of the ellipticity which drives $v_2\{2\}$ and projected $\vec{B}$ on participant plane as shown in MC-Glauber simulations~\cite{Chatterjee:2014sea}.

\begin{figure}[t]
 \includegraphics[width=1\textwidth]{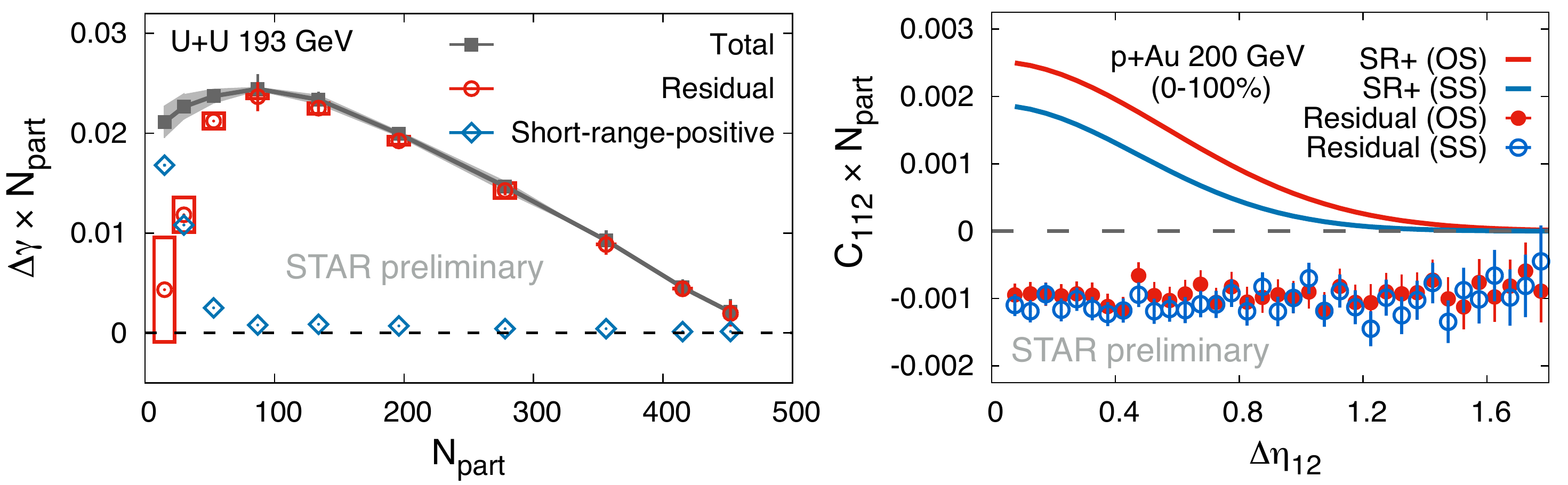}
%\hspace{2pc}%
\caption{\small \label{gammacent} (left) Centrality dependence of different components of $\Delta\gamma$ as defined in Eq.\ref{eq_c112}. The uncertainties from fits are shown in bars and the systematic uncertainties are shown by boxes (residual component) and band (total $\Delta\gamma$). (right) $\Delta\eta$ dependence of the three-particle correlator (numerator of the $\gamma$-correlator) for p+Au collisions showing the short-range-positive (SR$+$) and residual components in Eq.\ref{eq_c112}.}
\end{figure}

In Fig.\ref{gammav2} (right) we show the relative rapidity $\Delta\eta\!=\!\eta_1\!-\!\eta_2$ dependence of the charge inclusive three-particle correlator $C_{112} (\eta_1-\eta_2)=\langle \cos(\phi_1 (\eta_1)+\phi_2 (\eta_2)-2\phi_3)\rangle$ for a single centrality bin of $30-40\%$ in U+U collisions. %
One can see that over most of the range of $\Delta\eta$, $C_{112}$ remains negative. However, an interesting structure is observed at small $\Delta\eta$ where it trends towards positive values. It also changes sign at large $\Delta\eta$. Similar $\Delta\eta$ dependence is also observed for charge dependent correlators. Such trends might indicate different underlying phenomena driving the structure of $C_{112}(\Delta\eta)$~\cite{Adamczyk:2017hdl}. In particular, causality arguments~\cite{Dumitru:2010iy} indicate correlations at smaller $\Delta\eta$ are dominated by late time effects such as HBT, Coulomb, fragmentations {\it etc.}, whereas any early time effects that are driven by magnetic field or initial-state geometry can spread over large $\Delta\eta$. The effects of initial state geometry, hydrodynamic response and also momentum conservation have been previously shown to lead to negative values of $C_{112}\!<\!0$ (see Ref.~\cite{Bzdak:2010fd,Adamczyk:2017hdl,Teaney:2010vd}). 
On the other hand, one can show that in the short-range limit $(\Delta\eta, \Delta\phi \rightarrow0)$, $C_{112}\!>\!0$, 
indicating a possible tendency of the observable to become positive due to local clustering of particles caused by aforementioned late time effects that are the potential backgrounds for CME.  
%
%whereas 
%
 Based on such motivation we apply a data driven approach to separate different components of $C_{112}$ by fitting the $\Delta\eta$ distribution shown in Fig.\ref{gammav2} (right) with a function %al form of 
\begin{equation}
C_{112}(\Delta\eta_{_{12}})=A_{\!_{S\!R+}} e^{-(\Delta\eta)^2/2\sigma_{\!_{S\!R+}}^2} - A_{_{IR}} e^{-(\Delta\eta)^2/2\sigma_{IR}^2} + A_{_{LR}}.
\label{eq_c112}
\end{equation}
Here $A_{\!_{S\!R+}}$, $A_{_{IR}}$ are the amplitudes of the short-range positive and intermediate-range components respectively. The fit results indicate $\sigma_{\!_{S\!R+}}\le0.46\pm0.03$ and $\sigma_{\!_{IR}}\ge0.66\pm0.04$, {\it i.e.} a clear separation of widths exists between the two components. However the important distinction between the two components is that they differ in sign. 
The pedestal component $A_{_{LR}}$ is constant in $\Delta\eta$ and accounts for the sign change of $C_{112}$ at large $\Delta\eta$. A similar decomposition technique to remove short-range correlations in two-particle correlations has been recently used by STAR in Ref.~\cite{Adamczyk:2016exq}. In the current work, our goal is to study the charge dependence of different components in Eq.\ref{eq_c112}. For this we fit $C_{112}(\Delta\eta_{_{12}})$ separately for the same-sign and opposite-sign with Eq.~\ref{eq_c112} to extract the short-range positive ($=A_{\!_{S\!R+}} e^{-(\Delta\eta)^2/2\sigma_{\!_{S\!R+}}^2}$) and the residual components ($=C_{112} (\Delta\eta)-A_{\!_{S\!R+}} e^{-(\Delta\eta)^2/2\sigma_{\!_{S\!R+}}^2}$) in each centrality. We then estimate $\Delta\gamma = (C_{112}^{{opposite-sign}}-C_{112}^{same-sign})/v_{2}\{2\}$ separately for the two components. In order to account for of the trivial dilution effects while going from central to peripheral events, we present our results by multiplying $\Delta\gamma$ with $N_{part}$ in Fig.\ref{gammacent} (left). 

%\clearpage

Interestingly, in Fig.\ref{gammacent} (left) one can see that the residual component, which is wider in rapidity, shows a non-monotonic trend in contrast to the short-range positive component which increases with decreasing $N_{part}$. The total value of  $\Delta\gamma$ before the decomposition is also shown on the same plot. This indicates that for peripheral events $\Delta\gamma$ is dominated by short-range correlations. 
A similar dominance of short-range correlations was also seen in minimum bias p+Au collisions as demonstrated in Fig.\ref{gammacent} (right) where the difference between the residual components for both the same-sign and opposite-sign correlators seems to vanish. 
In Fig.\ref{gammacent}, the vanishing trend of the residual component in central, peripheral U+U and min-bias p+Au collisions can be explained by the disappearance of projected magnetic field in all these scenarios~\cite{Chatterjee:2014sea,Belmont:2016oqp}. The short-range positive component, on the other hand, seems to be a significant background source for charge separation that can lead to non-zero $\Delta\gamma$ in all such scenarios. 

\section{Summary}
In summary, we present STAR results on charge-dependent three-particle correlations for the search for the CME in A+A and p+A collisions at RHIC. In particular, we study the correlation between charge separation and elliptic flow, and the relative rapidity dependence of the correlation between the two particles carrying the charges. We find that the charge separation in peripheral A+A and in p+A collisions are dominated by short-range correlations. After removing such a short range positive component in each centrality bin, we find a residual component of the charge separation which is relatively wider in $\Delta\eta$ and vanishes towards central A+A, peripheral A+A and in min-bias p+A collisions. In such scenarios, MC-Glauber simulations predict strong decorrelation of $\vec{B}$ with the second harmonic event plane. 
In all such cases a naive background model, expected to predict non-vanishing charge separation, will be largely constrained. Going beyond naive background expectations, sophisticated theoretical inputs are much desired to see if the observed vanishing trends of charge separation are explained without invoking $\vec{B}$ driven effects. In the scenarios where the observed charge separation are large, {\it i.e.} in mid-central events, current measurements using $\gamma$-correlator do not provide discriminatory power to qualitatively disentangle the signal and background scenarios. Such limitations are results of the similar centrality dependence of flow and $\vec{B}$. Future isobar collisions at RHIC will provide better ways to disentangle the background and $\vec{B}$ driven effects.

%\vspace{-10pt}

%\section*{References}

%% References


\bibliographystyle{elsarticle-num}
\bibliography{qm2017_tribedy.bib}






% \end{thebibliography}

\end{document}